\documentclass[manuscript,screen]{acmart}

\usepackage{blindtext}
\usepackage{graphicx}
\usepackage{subcaption}
\usepackage[export]{adjustbox}
\usepackage{wrapfig}
\usepackage{hyperref}
\hypersetup{
    colorlinks=true,
    linkcolor=blue,
    filecolor=magenta,      
    urlcolor=cyan,
    pdftitle={Overleaf Example},
    pdfpagemode=FullScreen,
    }

\AtBeginDocument{%
  \providecommand\BibTeX{{%
    \normalfont B\kern-0.5em{\scshape i\kern-0.25em b}\kern-0.8em\TeX}}}

\begin{document}

\title{Virtual Takeovers in the Metaverse: Interrogating Power in Our Past and Future(s) with Multi-Layered Narratives}

\author{Heather Snyder Quinn}
%\orcid{}
\email{hquinn2@depaul.edu}
\affiliation{%
  \institution{DePaul University}
  \city{Chicago}
  \state{IL}
  \country{USA}
}

\author{Jessa Dickinson}
%\orcid{}
\email{jessa@idc.coop}
\affiliation{%
  \institution{DePaul University}
  \city{Chicago}
  \state{IL}
  \country{USA}
}


\begin{abstract}   
\textit{Mariah} is an augmented reality (AR) mobile application that exposes power structures (e.g., capitalism, patriarchy, white supremacy) through storytelling and celebrates acts of resistance against them. People can use \textit{Mariah} to ``legally trespass'' the metaverse as a form of protest. \textit{Mariah} provides historical context to the user’s physical surroundings by superimposing images and playing stories about people who have experienced, and resisted, injustice. We share two implementations of \textit{Mariah} that raise questions about free speech and property rights in the metaverse: a protest against museums accepting ``dirty money’’ from the opioid epidemic, and a commemoration of sites where people have resisted power structures. \textit{Mariah} is a case study for how experimenting with a technology in non-sanctioned ways (i.e., ``hacking'') can expose ways that it might interact with, and potentially amplify, existing power structures.

\end{abstract}




\begin{teaserfigure}
  \includegraphics[width=\textwidth]{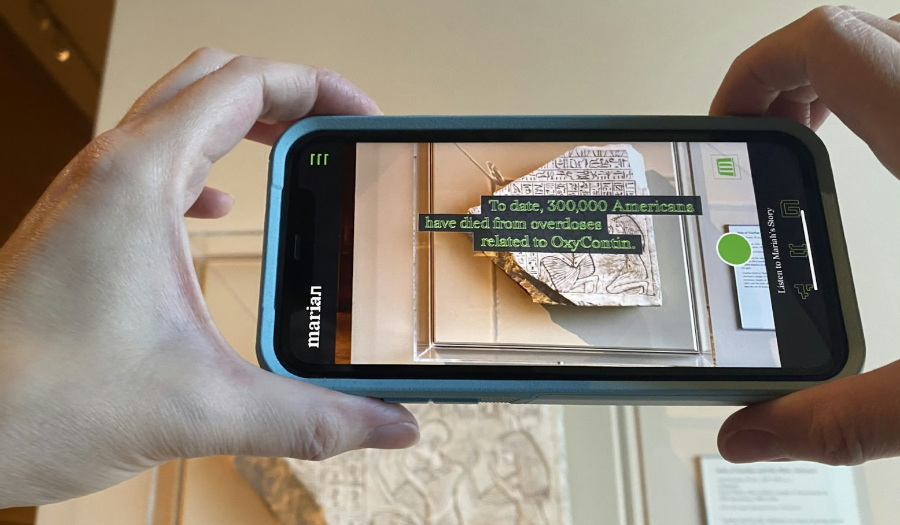}
  \caption{\textit{Mariah} augments an artifact in the Metropolitan Museum of Art's Dendur Wing with a statistic about the opioid crisis.}
  \Description{Image of Mariah overlay at the New York Met, statistic reads: To date, 300,000 Americans have died from overdoses related to OxyContin.}
  \label{fig:teaser}
\end{teaserfigure}


\maketitle

\section{Introduction}
People create and use technologies in the context of interconnected structures of power (e.g., capitalism, white supremacy, patriarchy \cite{collins2019critical, crenshaw1990mapping}). Traditional companies are governed by capitalist rules, such as legally requiring board members to protect and maximize the company's profits. For instance, board members of Purdue Pharma cited their feduciary duties in defense of their production and aggressive marketing of a highly addictive pain killer and opioid called OxyContin \cite{NYT2019SacklersLawsuits}. The Sackler family and their company, Purdue Pharma, caused the opioid epidemic in the United States with their perverse strategies to addict people to their drug and then sell them addiction treatment \cite{NYT2019SacklersLawsuits, alpert2022originsOpioid}. From 1997 (when OxyContin was released) to 2017, deaths caused by drug overdoses in the United States increased by 3.6\%  \cite{NCHS99-2017overdoses}. The epidemic has claimed hundreds of thousands of lives \cite{NPR2023SacklersSupremeCourt}, and has exasperated preexisting inequities along the fault lines of race, class, gender, and region \cite{hansen2016opioidWhite, singh2019opioidSocial}. 

The first author and her collaborators designed \href{https://apps.apple.com/us/app/mariah/id1628118349
}{\textit{Mariah}}, an augmented reality (AR) application (app), to expose the power structures behind the opioid crisis. They used \textit{Mariah} to critique the Metropolitan Museum of Art (the ``Met'') and the Louvre for accepting ``dirty'' donations from the Sackler family \cite{NYTSacklerMet}. The team also implemented \textit{Mariah} in San Francisco as part of an art installation to tell the stories of historic protests at different sites across the city. \textit{Mariah} creates a guerilla-style AR experience in which users ``legally trespass'' the metaverse, viewing and listening to content that tells a narrative that adds complexity and historical context to their physical surroundings. We use \textit{Mariah} as a case through which to explore how using technologies in subversive ways can reveal potential future impacts the technology could have on society. We ask:  

\begin{enumerate}
    \item \textit{How might augmented reality be used to create counter-narratives that expose and challenge power structures?} 

    \item \textit{What are the legal challenges and questions that arise from exploring the metaverse as a site of protest?} 
\end{enumerate}

We contribute to literature in human-computer interaction (HCI) that considers how social and institutional contexts shape how technologies are designed, governed, and used---as well as the impacts they have on society \cite{toyama2015, thebault-spieker15}. Drawing from afrofuturist speculative design \cite{Harrington2022Futuring, harrington2021speculativeCodesign, AntiRacistSpecDesSigchi, benjamin2019captivating, bray2022radicalFutures}, we frame \textit{Mariah} as a tool with which to explore how power structures (and technologies) might shape our potential futures  \cite{collins2019critical}. For instance, a collaborative team of community members, designers, lawyers, and policymakers could use \textit{Mariah} as a probe \cite{mattelmaki2006probes, Tomlinson202accountableCapitalism} to consider preposterous, plausible, and probable futures. Or, perhaps more importantly, liberatory futures. This continues the work in HCI that interrogates structures such as capitalism (e.g., \cite{Feltwell2019PostCapitalism, Tomlinson202accountableCapitalism}), white supremacy (e.g., \cite{ogbonnaya2020crt, Erete2020breathe, To22interactive}), and colonialism (e.g., \cite{irani2010postcolonial, Das2022decolonizeNarrativeAgency}).

\section{Related Work}

\textit{Mariah} uses storytelling as a vehicle to personalize injustice and encourage reflection in ways that traditional media and statistics may not, as found by Zhang, et al. (2021) \cite{Zhang21nudgeReflection}. Drawing from \cite{solorzano2002counterstory}, we explore how we can use counter-narratives in our work as a way to examine \cite{Silva22ARactivism, rankin2021saturated}, cope with \cite{To22interactive}, and resist \cite{Das2022decolonizeNarrativeAgency, dickinson2021counter} dominating power structures. Our focus on structural power follows work that takes a transformative justice approach to expand our HCI approaches to issues to include the structural factors that create them \cite{erete2021TJ, HawraRabaan2021TJmuslim, Musgrave2022TJblackWmnJoy}. Continuing the agenda set out by Silva, et al. (2022) \cite{Silva22ARactivism}, we raise concerns about how free speech, property rights, and privacy in AR will be addressed in the context of capitalism and legislative structures that are rooted in racial capitalism \cite{gilmore2007golden, robinson2020racialcap, melamed2015racialcap}.

\section{Augmented Reality: The Basics}

Augmented reality (AR) is an emerging technology that superimposes a computer-generated image on a user’s view of the real world, thus providing a composite view. AR is one of the ways in which users see and access the metaverse (also called the AR cloud and the spatial web). Currently, AR can only be viewed via a limited number of applications (apps), and there is a risk that the spatial web will become dominated by a centralized authority. The potential for monopoly control of the metaverse has raised alarm in the HCI community \cite{Silva22ARactivism}. In order to discuss the social implications of AR, we share a basic description of how the technology works. AR can currently be implemented in three ways:

\begin{enumerate}

    \item Trigger Image: An AR application uses a phone’s camera to scan for a particular image (e.g., a QR code). When the application recognizes the image, it will reveal an image, text, video, or sound to the user.

    \item Geolocation: A phone's latitude and longitude coordinates can be used as an AR trigger. When a user enters the trigger area, the app will augment the physical environment. 

    \item Ground/Plane: AR can virtually place an object in a user's environment using their camera display. 

\end{enumerate}

Mariah uses all of these technologies to superimpose images and text on the user's display, and to play audio of stories relevant to the surroundings. The first author and her collaborators created Mariah using public photographs as the trigger images. They did not have to set foot in the museum to create an application that could augment it, which raised ethical and legal questions we will explore in this paper.
\section{Mariah}

\textit{Mariah} is a collaboratively developed initiative that challenges institutional power, including big tech and big pharma, by ``hacking the metaverse'' as an act of protest. The application is named after Mariah Lotti, who lost her life to opioids at 19 years old. The first author created the initiative in partnership with Mariah's mother, Rhonda Lotti, and Adam DelMarcelle, who lost his brother to opioids. They are designers and artists and approached \textit{Mariah} as a protest and a way to personally process the premature deaths of their loved ones. Their second implementation (\textit{``Strikethrough''}) of Mariah expanded its scope to include the commemoration of sites of protest. In this section, we describe both implementations of \textit{Mariah}. For more information, see the short \href{https://filmfreeway.com/MariahActsofResistance
}{documentary film} \textit{Mariah} (which premiered in 2023 at \textit{The New Media Film Festival} and won Best AR Film), a \href{https://drive.google.com/file/d/1TMEbXrIGu6PZqGS_yKxhhHCU1KsN_yyx/view?usp=drive_link
}{video demonstration}  of the \textit{Mariah} at the Met, and the \href{https://www.instagram.com/reel/CqCswKvNdJu/?hl=en
}{\textit{Strikethrough}} Instagram reel, and press articles \cite{Mariah2020Post, Mariah2020Hyperallergic}.

\subsection{Exposing ``Dirty Money'' in Museums}

The team initially designed \textit{Mariah} for use in the Temple of Dendur in the Met. At the time, the wing of the museum that holds the iconic Dendur Temple was named for the Sackler family, in honor of their financial contribution to the museum (their name has since been removed) \cite{NYTSacklerMet}. Launched in 2020, \textit{Mariah} transformed the Met’s Sackler wing into a virtual memorial for Mariah Lotti and others who had lost their lives to the opioid epidemic. Exploring AR’s potential to revise historical narratives and its ethical implications, the app augments Sackler donated art with ``virtual memorials'' (audio and video of the lives of opioid victims). \textit{Mariah} is a publicly available mobile app. The first author and her team did not gather data on app usage because they created it as a protest and memorial, rather than a research study. People learned about \textit{Mariah} through word of mouth and media coverage.

The augmented experience includes audio, video, and graphics that correspond with sixteen (16) Sackler-donated artifacts in the Dendur Wing. Each sculpture is augmented visually with a global statistic about the opioid epidemic. The app also plays audio of Mariah’s mother telling stories about Mariah’s life to humanize the opioid epidemic. The experience concludes in the Temple of Dendur, where the viewer is able to place a virtual memorial in the room that renames the room ``The Mariah Lotti Memorial Gallery” (see figure 2). 

Following the ``installation'' of \textit{Mariah} at the Met, the artists expanded the project to The Pyramid du Louvre in Paris, France (which also received Sackler donations \cite{NYTSacklerMet}) with a geolocative installation. Titled ``Funded by Loss: Death in Real-Time,'' the installation is a large-scale, 3-dimensional number that increases every 5 minutes to represent real-time opioid fatalities (see figure 2). This installation also tells Joseph DelMarelle's story, who lost his son to an epidemic that was created by, and enriched, the Sackler family and Purdue Pharma \cite{alpert2022originsOpioid, NYTSacklerMet}.

\begin{figure}
  \includegraphics[width=\textwidth]{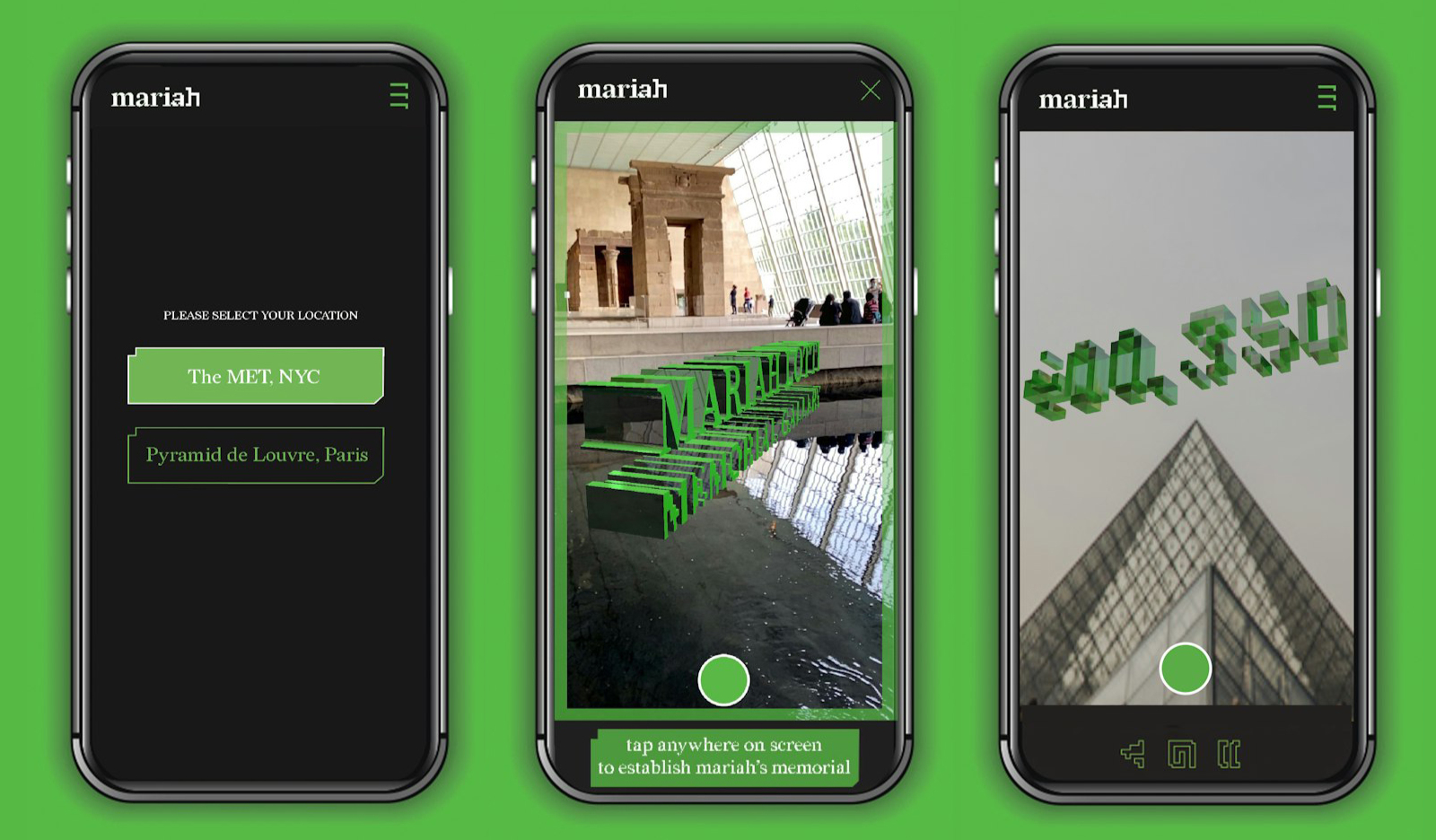}
  \caption{Screenshots of the museum installations of \textit{Mariah}. The middle screen shows the virtual memorial for Mariah Lotti ``placed'' in the Temple of Dendur in the Met. The right screen shows the opioid ``death clock'' superimposed over the Louvre.}
  \Description{}
  \label{fig:teaser}
\end{figure}

\subsection{Commemorating Sites of Protest with Mariah}

In 2022, the Letterform Archive commissioned the first author and her team to expand the Mariah app as a vehicle of protest for the \textit{Strikethrough: Typographic Messages of Protest} exhibition. They used \textit{Mariah} to augment seven sites of historic protest with historic audio and video throughout the San Francisco Bay area.  The locations were marked on a map and with removable spray chalk on sidewalks (see figure 3).  The sites included:

\begin{itemize}
    \item BLM: Black Lives Matter march across the Golden Gate Bridge, 2020

    \item IOAT: Indians of all tribes, Occupation of Alcatrez, 1969-71

    \item MARIAH: War on drugs at Civic Center and the Tenderloin, 2022

    \item MILK: Harvey Milk memorial march on Castro and Market Streets, 1979

    \item STRIKE: Shipyard worker strikes near the Bethlehem Steel building, 1941

    \item FREE SPEECH: Student protests and poster printing on the Berkeley campus, 1960s

    \item PANTHERS: Central HQ of the Black Panthers in West Oakland, 1967-71

    \item OCCUPY: Shutdown of Oakland Port by Occupy movement for economic equality, 2011
\end{itemize}

\begin{figure}
  \includegraphics[width=\textwidth]{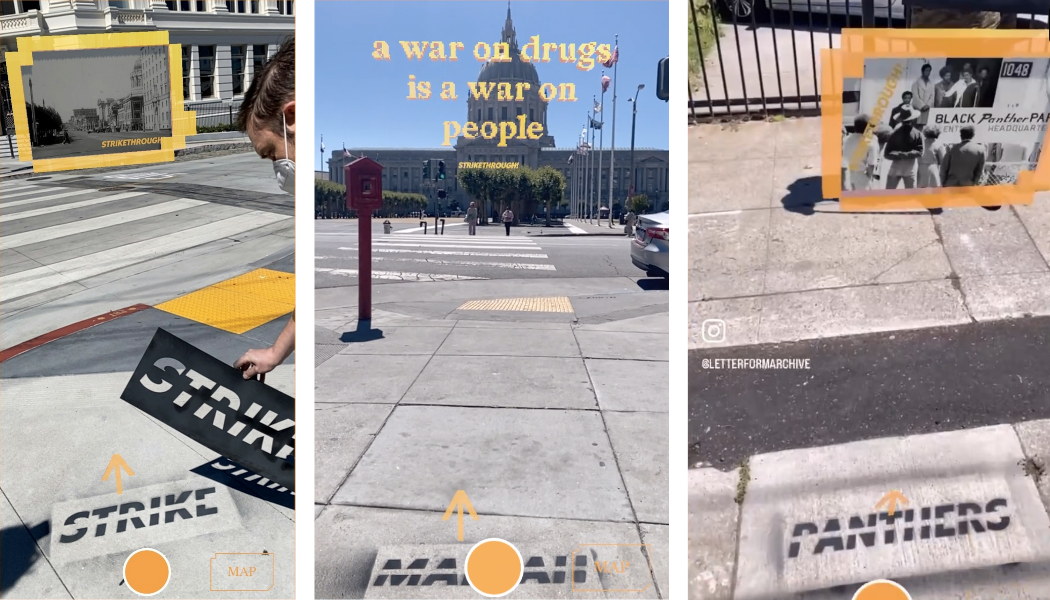}
  \caption{Three locations where \textit{Mariah} was implemented for \textit{Strikethrough}. From left to right: Strike, commemorating labor protests; Mariah, protesting the war on drugs and the opioid epidemic; and Panthers, commemorating the Black Panther's headquarters.}
  \Description{}
  \label{fig:teaser}
\end{figure}

\section{Discussion}

\subsection{Complicating Narratives and Exposing Power with AR}
The museum and \textit{Strikethrough} installations of \textit{Mariah} addressed different topics, but both used narratives to reveal power structures and complicate narratives about societal issues. \textit{Mariah} shifts the narrative focus of the the opioid epidemic and the ``war on drugs'' from the individuals suffering from addition, to the people and institutions that caused these crises \cite{Aizen2018OpioidResponse, NYT2019SacklersLawsuits}. The personal stories told through \textit{Mariah} also serve to humanize the people lost in this epidemic, which resists narratives that dehumanize opioid victims (e.g., referring to people as ``addicts,'' ``drug users''). In this way, \textit{Mariah} seeks to build a counter-story \cite{solorzano2002counterstory} about the opioid epidemic and the war on drugs.  \textit{Strikethrough} also leveraged \textit{Mariah} as a tool to celebrate collective action against capitalism, white supremacy, and cis-hetero-patriarchy by creating virtual memorials and sharing historical stories.

\textit{Mariah} was a project that was rooted in personal experience and loss. The people who created \textit{Mariah} worked from their experiences of the opioid crisis as white people, and shared the personal stories of their loved ones from that standpoint. In the museum installations of \textit{Mariah}, they did not take on the intersectional \cite{collins2019critical} dimensions of the opioid epidemic \cite{hansen2016opioidWhite, singh2019opioidSocial}
. The \textit{Strikethrough} installation gave them the opportunity to examine and protest the institutions and power structures that created the racist war on drugs beginning in the 1970s and subsequent concentrated incarceration of Black and Brown neighborhoods \cite{alexander12massIncarceration, lugalia2018war, travis2014incarceration}. The opioid epidemic that began in the 1990s has been treated more so as a public health issue than a criminal issue, largely due to the racial and class differences between the populations most impacted \cite{hansen2016opioidWhite}.

\textit{Mariah} is an example of how AR can be leveraged as a tool for examining power and building counter-narratives. Although the Met was not directly involved in the opioid crisis, they benefited from it in the form of large donations from the Sacklers. The Met is a powerful cultural institution held in high regard by the public, and \textit{Mariah} complicated the public narrative about the museum while prompting reflection on the impacts of capitalism on our society. Power structures such as capitalism are codified and implemented through our legal system, which will be tasked with governing the metaverse in the future.

\subsection{Legal Questions in the Metaverse}
The artists who created \textit{Mariah} did not receive permission to augment the Met, but consulted with an attorney and received a letter of support regarding its legality under the First Amendment. The experience of ``installing'' \textit{Mariah} prompted the first author to consider future implications of our human freedoms in the metaverse, including interpretations of free speech and property rights (both physical and intellectual). Who owns our virtual space and what can be placed there? According to lawyer Brian Wassom, ``of particular concern is how AR will challenge traditional property rights and stretch interpretations of free speech.'' Wassom asserts that AR content is an expression akin to digital speech and is protected by the First Amendment. If this is correct, then the digital space around (maybe even inside) your property could become ``a kind of common land making every street, every surface, every portion of space with a GPS coordinate, free real estate for those with the ability and the will to augment.'' The implications are concerning, especially considering that the same institutions and power structures that created the war on drugs and the opioid epidemic will be tasked with governing the metaverse. We pose that the method the first author employed with \textit{Mariah} (i.e., using AR in an experimental, subversive way) can be used to as a speculative design probe \cite{Harrington2022Futuring, AntiRacistSpecDesSigchi, bray2022radicalFutures, mattelmaki2006probes} to explore technologies' potential societal impacts and collectively envision more just futures.

\section{Conclusion}

The Met never commented on \textit{Mariah}, though it replaced some artifacts in the Dendur Wing, which the artists re-augmented. Following litigation against the Sacklers and Purdue Pharma \cite{NYT2019SacklersLawsuits}, the Met removed the Sackler name \cite{NYTSacklerMet}. Although this is a positive step, simply removing the Sackler name obscures the complex history of funding and harm instead of engaging with it. In future work, the first author will use AR to explore how museums have come to possess certain cultural artifacts, and the role that colonization and other forms of power played in the artifacts' journeys. 
\section{Acknowledgements}
Artists: Adam DelMarcelle, Heather Snyder Quinn. Curators: Silas Munro, Stephen Coles. Contributors: Mariah Lotti, Rhonda Lotti, Westfield Productions
Design and tech: Zishou Wang, Valerie Shur, Leslie Ramirez, Les Garcia, Jake Juraka, Flor Salatino. Funders: The Letterform Archive, Washington University in St Louis, DePaul University. The Mariah app and website use \href{https://www.redaction.us/
}{\textit{Redaction}}—a typeface that charts the history and abuses of the US criminal justice system.

\bibliographystyle{ACM-Reference-Format}
\bibliography{references}

\end{document}